# A Systematic Mapping Study and Practitioner Insights on the Use of Software Engineering Practices to Develop MVPs

Silvio Alonso, Marcos Kalinowski, Bruna Ferreira,
Simone D. J. Barbosa, Hélio Lopes

*Department of Informatics - Pontifical Catholic University of Rio de Janeiro (PUC-Rio), Rio de Janeiro, Brazil*

---

## Abstract

**[Background]** Many startup environments and even traditional software companies have embraced the use of MVPs (Minimum Viable Products) to allow quickly experimenting solution options. The MVP concept has influenced the way in which development teams apply Software Engineering (SE) practices. However, the overall understanding of this influence of MVPs on SE practices is still poor. **[Objective]** Our goal is to characterize the publication landscape on practices that have been used in the context of software MVPs and to gather practitioner insights on the identified practices. **[Method]** We conducted a systematic mapping study using a hybrid search strategy that consists of a database search and parallel forward and backward snowballing. Thereafter, we discussed the mapping study results in two focus groups sessions involving twelve industry practitioners that extensively use MVPs in their projects to capture their perceptions on the findings of the mapping study. **[Results]** We identified 33 papers published between 2013 and 2020. We observed some trends related to MVP ideation (or MVP conception) and evaluation practices. For instance, regarding ideation, we found six different approaches (*e.g.*, Design Thinking, Lean Inception) and mainly informal end-user involvement practices (*e.g.*, workshops, interviews). Regarding evaluation, there is an emphasis on end-user validations based on practices such as usability tests, A/B testing, and usage data analysis. However, there is still limited research related to MVP technical feasibility assessment and effort estimation. Practitioners of the focus group sessions reinforced the confidence in our results regarding ideation and evaluation practices, being aware of most of the identified practices. They also reported how they deal with the technical feasibility assessments (involving developers during the ideation and conducting informal experiments) and effort estimation in practice (based on expert opinion and using practices common to agile methodologies, such as Planning Poker). **[Conclusion]** Our analysis suggests that there are opportunities for solution proposals and evaluation studies to address literature gaps concerning technical feasibility assessment and effort estimation. Overall, more effort needs to be invested into empirically evaluating the existing MVP-related practices.

---

## 1. Introduction

In Lean Startup (Ries, 2011), Eric Ries presented a methodology for allowing entrepreneurs to develop new products using validated learning about the customer. What initially began to be used in startups has gained popularity and nowadays many software companies are adopting parts of this methodology to achieve high levels of innovation. Its main focus relies on the identification and implementation of a product that adds real value to the customer (Paternoster et al., 2014). Indeed, a key concept to Lean Startup is the Minimum Viable Product (MVP).

A recent systematic mapping about the term MVP (Lenarduzzi and Taibi, 2016) showed that there are many definitions; the most popular one was provided by Ries, who defines an MVP as "a version of a new product, which allows a team to collect the maximum amount of validated learning about customers with the least effort". That mapping also reported that the key factors related to MVPs are technical characteristics of the product and market and customer aspects, such as "minimum set of features", "customer feedback", and "minimum effort" (Lenarduzzi and Taibi, 2016).

Lean Startup advocates for building MVPs as experiments to perform the "build-measure-learn" cycle as fast as possible (Ries, 2011). MVPs are being used in many different contexts, such as: startups (Duc and Abrahamsson, 2016), universities (Mardhia and Anggraini, 2019), industry-academia collaborations (Münch et al., 2013; Kalinowski et al., 2020), and established enterprises (Dennehy et al., 2019; Edison et al., 2018). Nevertheless, there is not much evidence on how this use is affecting software engineering practices. Lindgren and Münch (2016) found that, even though the principles of continuous experimentation resonated with industry practitioners, the state of practice is not yet mature. Bridging the gap between the theory in software engineering and the practical use of MVPs is a relevant issue that will benefit both practitioners from industry and academia.

Considering this scenario, we synthesized existing work on practices em-ployed in the development of software products using MVPs (Alonso et al., 2021). We reported on a secondary study yielding 33 selected publications to pursue our research objective that is to understand Software Engineering practices that have been used to develop software product MVPs. We characterized these practices by answering six Research Questions (RQs). RQs 1–4 are related to four software MVP development activities, respectively: ideation (also known as conception), technical feasibility assessment, effort estimation, and evaluation. RQs 5–6 were related to characterizing the reported evidence. The RQs are detailed in Section 3.

In this paper we extend our previous effort (Alonso et al., 2021) by providing more details on the mapping study and also presenting results of two additional focus group sessions conducted to discuss the findings of the systematic mapping related to RQs 1–4. These sessions were attended by professionals active in innovation projects within an industry-academia collaboration initiative, in which the MVP concept has been extensively applied.

Our results show that most papers concern practices regarding MVP ideation and evaluation. For ideation, we found the use of several approaches (*e.g.*, Design Thinking, Lean Inception) and an emphasis on informal end-user involvement practices (*e.g.*, workshops, interviews) and lightweight documentation (*e.g.*, features, user stories). Regarding evaluation, there is an emphasis on end-user validations based on different practices (*e.g.*, usability tests, A/B testing). The focus groups confirmed these findings, as the participants were aware of most of these practices and had been applying some of them as part of their projects.

We also identified literature gaps concerning technical feasibility assess-ment and effort estimation, indicating a need for further empirical studies. Analyzing the discussions held during the focus group sessions, we identified practices used within the participants' projects regarding technical feasibility assessment and effort estimation. Industry practitioners reported involving technical experts during the ideation process and conducting informal experiments as ways of assessing technical feasibility. Regarding effort estimation, practitioners relied on expert opinion and practices common to agile methodologies, such as Planning Poker.

The main contribution of this paper consists in providing a comprehensive literature review on software engineering practices used to develop MVPs and complementing it with practitioner insights. Besides allowing to understand which practices are being used, we highlight relevant gaps to be addressed, e.g., the lack of solution proposals concerning MVP technical feasibility and effort estimation.

The remainder of this paper is organized as follows. Section 2 presents the background and related work. Section 3 presents the overall goal of our research and our research questions. Section 4 defines the systematic mapping protocol, whereas Section 5 defines the focus group protocol. Section 6 presents the results, which are further discussed in Section 7. Finally, Section 8 concludes the paper.

## 2. Background and Related Work

In this section, we outline the domain focusing on software engineering and the role of MVPs. Related work concerns other secondary studies conducted on closely related topics.

### *2.1. Software Engineering and MVPs*

Software engineering is concerned with all aspects of software production, from the early stages of system specification to maintaining the system after it has gone into use (Sommerville, 2016). As an engineering discipline, it focuses on applying the right theories, methods, and tools to get results with the required quality within the proposed schedule and budget.

The need for product development methodologies to be adapted to the scope, size, complexity, and changing requirements in the initial phase of a software project is widely recognized. However, there is still little guidance on how startups and innovation companies can adapt their process to rapid changes in context (Pantiuchina et al., 2017). Many innovation companies carry out problem-solving experiments to learn more about different solution options, often leading to delivering a product that is very different from the original idea (Nguyen-Duc et al., 2017). This learning is typically associated with the construction of a Minimum Viable Product (MVP), a lean version of the product to validate a new technology or elicit customer requirements (Nguyen-Duc et al., 2017; Ries, 2011).

Ries (2011) proposed the Lean Startup methodology for business and product development. He/She defined a product creation process that can be summarized by combining business-oriented hypothesis experimentation and iterative product launches. Building a product iteratively, taking into account the needs of initial customers, can reduce risks such as expensive launches, failure to use, and low adherence. In Lean Startup, Build-Measure-Learn is the fundamental principle to transform ideas into products, measure how customers respond, and, finally, know whether to give up or persevere (Ries, 2011). A fundamental concept of Build-Measure-Learn is building MVPs based on quick feedback obtained from the initial users. Ries (2011) defined an MVP as: ”a version of a new product, which allows a team to collect the maximum amount of validated learning from customers with minimum effort.”

Several other definitions have been proposed (Lenarduzzi and Taibi, 2016), and practitioners and researchers often face the problem of selecting the most appropriate one. Moreover, the influence of MVPs on software engineering practices is still poorly understood.

### *2.2. Related Work*

To the best of our knowledge, only one systematic mapping study was conducted focusing solely on MVPs, but we identified papers that present secondary studies concerning Lean Startup and MVP.

Lenarduzzi and Taibi (2016) presented a mapping study about the different definitions of MVPs in the literature. They found 22 papers, proposed a classification schema for characterizing the definition of MVP in Lean Startups, and identified a set of common key factors in the MVP definitions. While we also focus on MVPs and build on their definitions, our study has a different scope, as we aim at characterizing practices that are being used in the context of MVP development.

Paternoster et al. (2014) conducted a systematic literature review on software development work practices in startup companies. They identified 43 primary studies, of which only 16 were entirely dedicated to software development in startups. Moreover, only 9 studies exhibited high scientific rigor and relevance. The authors did not focus their discussions on the use of MVPs in the development of software systems.

Berg et al. (2018) presented a systematic mapping on startup research from an engineering perspective, involving 27 papers published before 2017. They aimed to identify thematic concepts involved in startup research.

It is noteworthy that only the first of these three studies focuses directly on MVPs, but was not scoped on mapping software engineering practices. The second study reviews development practices for startups. Although startups are known as great adopters of MVP practices, the findings cannot be generalized to the MVP context. The third one focused on thematic concepts involved in startup research. The widespread adoption of MVPs in different contexts deserves specific attention to better understand the various practices, which is the focus of this paper.

## 3. Goal and Research Questions

The main goal of our research is to **characterize software engineering practices that have been applied to develop software MVPs.** Our research questions focus primarily on the fundamental activities of specification and validation.

*RQ1. Which practices have been used to ideate MVPs?* We aim to provide a clear view of one of the first activities of an MVP life cycle. Comparable to typical initial requirements elicitation efforts of traditional software products, ideation aims at defining the desired features of the MVP, outlining what the MVP will provide to the customer.

*RQ2. Which practices have been used to assess the technical feasibility of MVPs?* It is important to assure that the desired MVP features are technically feasible, *i.e.*, can be built. This task is somewhat comparable to assessments made during typical software engineering elaboration phases. Failing to assess the technical feasibility of any software project during its early stages might imply a waste of investment and commitment to initiatives that are doomed to fail.

*RQ3. Which practices have been used to estimate the MVP building effort?* We aim to provide an overview of practices that are employed to estimate MVP effort. As for any software product, from a business perspective, it is important to know how much effort an MVP requires to be built.

*RQ4. Which practices have been used to evaluate MVPs?* One of the main goals of an MVP is to collect validated learning from the customer. To this end, the evaluation of an MVP should collect data to generate insights from the customer's perspective. In software engineering, this is akin to continuous experimentation (*e.g.*, A/B Testing) (Fitzgerald and Stol, 2017).

*RQ5. Which type of research has been conducted regarding MVPs?* To classify the types of research identified, we adopted the scheme proposed by Wieringa et al. (2006): evaluation research, solution proposal, validation research, philosophical paper, opinion paper, and experience paper.

*RQ6. Which types of empirical evaluation have been performed in research regarding MVPs?* We aim to identify the types of empirical evaluation (*e.g.*, case study, controlled experiment, survey) that have been performed, especially in evaluation or validation research papers.

To address RQs1-6 we conducted a systematic mapping study. To complement our findings regarding RQs1-4 with discussions from a practical perspective, we conducted two focus group sessions.

## 4. Systematic Mapping Protocol

A systematic mapping study is a form of secondary study that provides a systematic procedure and structure of the type of research reports and results that have been published, aiming to answer a particular research question (Petersen et al., 2015). This section presents our systematic mapping protocol, detailing the search strategy, study selection, and data extraction. The results of the research questions, complemented by discussions based on our focus group sessions, will be presented in Section 6.

### *4.1. Search Strategy*

A study comparing different search strategies to perform Systematic Literature Reviews (SLRs) in software engineering (Mourão et al., 2020) found that using a hybrid strategy combining a database search on Scopus with parallel backward and forward snowballing (using Google Scholar) tends to present an appropriate balance between result quality and review effort.

With this in mind, we adopted such hybrid search strategy, involving applying a search string on the Scopus database, filtering the results using our defined exclusion criteria, and performing parallel backward and forward snowballing iterations on the remaining studies (Mourão et al., 2017, 2020). By following this parallel process, papers obtained by backward snowballing were not subject to forward snowballing, and vice-versa.

To design our search string for the database search, we used the PICO criteria (Kitchenham and Charters, 2007), as follows. Population (P): Software; Intervention (I): Minimal Viable Product or MVP; Comparison (C): N/A; Outcome (O): N/A. Thus, our resulting search string was "'software'

AND (‘Minimum Viable Product’ OR ‘MVP’)”. It was applied it to titles, abstracts and keywords in Scopus in March 2021.

### *4.2. Study Selection*

Our study comprised papers published by the end of 2020. The study’s Inclusion Criterion (IC) and Exclusion Criteria (ECs) to filter the studies are presented in Table 1.

Table 1: Inclusion and Exclusion criteria

| Criteria | Description |
|---|---|
| IC1 | Describe practices related to the development of software MVPs |
| EC1 | Just mention MVPs but do not satisfactory describe the used practices |
| EC2 | Do not mainly comprise software development-related MVPs (*e.g.*, mainly hardware or IoT related MVPs) |
| EC3 | Not written in English |
| EC4 | Not published in a peer-reviewed conference, journal, or workshop |
| EC5 | Published after 2020 |

Our study selection process was performed by one researcher and revised by another. Whenever they disagreed or were in doubt on whether to include a paper, a third researcher was involved to reach consensus.

The search in Scopus returned a total of 223 studies. First, we applied the selection criteria to the title and abstract of all candidate studies, filtering papers that were barely related to the topic. Then, we applied the selection criteria to the full text, filtering them to obtain our seed set, formed by 27 studies.

The seed set was used as input to a parallel forward and backward snowballing. In the first backward snowballing iteration, we analyzed 496 papers and included three; in the second iteration, we analyzed 110 papers but included none. In the first forward snowballing iteration, we analyzed 313 papers and included two; in the second iteration, we analyzed five papers but included none. Together, the search and snowballing procedure resulted in 32 papers. The entire procedure is detailed in our online repository.[1]

The authors were aware of one additional study that was not retrieved by the search strategy (Kalinowski et al., 2020) and analyzed the reason for missing it. Although it concerned software-based MVPs and was indexed in Scopus, the paper did not use the term "software" in its abstract. Moreover, it was published in late 2020, so it was not a candidate for inclusion through backward snowballing. Finally, while the paper reported case studies with practices applied to MVP development, it had a broader scope and therefore did not refer to specific MVP research literature, also preventing its inclusion through forward snowballing. That paper was included manually. It is noteworthy that there may be other papers that we might have missed for similar reasons. Still, our analysis indicates that this one missed paper represents a very specific and justifiable case. Therefore, we believe that our final set of included papers comprises a representative sample to provide comprehensive and meaningful answers to our research questions.

In total, 33 papers were included in this study. An overview of these papers, organized by research type ("PP" stands for "philosophical paper") and year, and a short description of what the paper concerns is presented in Table II. The complete references of the included studies (S1-S33) can be found in Appendix A.

[1]https://doi.org/10.5281/zenodo.4718759

Table 2: Paper Descriptions Organized by Research Type and Year

| Type | Year | ID | Description |
|---|---|---|---|
| Evaluation Research | 2020 | S5 | Comparison between Scrum and Lean Inception in the initiation phase of a software project |
| | | S25 | Evaluation of interdisciplinary scenario-building |
| | | S28 | Analysis of challenges related to the steps of an MVP development |
| | 2019 | S6 | Evaluation of the effects of elements from a startup ecosystem on MVPs |
| | | S10 | Practices to perform requirements gathering in software startups |
| | 2018 | S2 | Investigation on how Lean internal startup facilitates software product innovation in large companies |
| | | S14 | Adaptation of Scrum to a product innovation context |
| | | S15 | Analysis of the relationship between hypotheses and MVPs in startups |
| | | S32 | Investigation of pivots from a Lean startup perspective |
| | 2017 | S1 | Investigation on how software startups approach experimentation |
| | | S20 | Investigation of factors that influence the speed of prototyping in software startups |
| | | S29 | Approach to perform effort estimation for change requests |
| | | S31 | Approach to perform effort estimation for change requests |
| | 2016 | S22 | Analysis of the MVP role in software startups |
| | 2013 | S24 | Approach to create MVPs in industry-academia collaborations |
| Personal Experience | 2020 | S3 | Development of a conversational agent (chatbot) |
| | | S4 | Lean UX development in a fintech context |
| | | S19 | Development of a digital tool for health care |
| | | S30 | Development of an app to support dental care of deaf people |
| | 2019 | S7 | Development process of an MVP to assist firefighters |
| | | S8 | Lean method contribution to a User Experience testing experiment in an academic context |
| | | S11 | Development of an MVP in a healthcare environment |
| | | S26 | Approach to streamline the requirements engineering process of mobile applications |
| PP | 2015 | S23 | Method to involve users to gain meaningful feedback and learning |
| Proposal of Solution | 2020 | S33 | Lean R&D approach for digital transformation |
| | 2019 | S9 | Approach to develop MVPs in established companies |
| | | S12 | Method to develop MVPs in software startups |
| | 2018 | S27 | Analysis of a customer touchpoint model implementation in a software development process |
| | 2017 | S16 | Approach to analyze user feedback on MVPs of mobile applications |
| | | S17 | Approach to adopt agile development on MVPs of mobile applications |
| | | S18 | Approach to analyze user feedback on MVPs of mobile applications |
| | | S21 | Approach to analyze user feedback on MVPs of mobile applications |
| | 2016 | S13 | Approach to analyze user feedback on MVPs of mobile applications |

### *4.3. Data Extraction*

Data extracted from each paper involved general publication metadata fields (*e.g.*, year, authors), the answers to each research question, and the advantages and disadvantages reported for some of the practices. Table 3 shows the data extraction form. The form and all extracted data are also available as a spreadsheet in our online repository.[1]

Table 3: Data Extraction Form

| **Information** | **Description** |
|---|---|
| Study metadata | Name, Authors, and Year of publication |
| Practices used to ideate (RQ1) | Name or short description of the practices |
| Advantages and disadvantages | Advantages and disadvantages reported for the ideation practices |
| Technical feasibility assessment practices (RQ2) | Name or short description of the practices |
| Advantages and disadvantages | Advantages and disadvantages reported for the technical feasibility assessment practices |
| Effort estimation practices (RQ3) | Name or short description of the practices |
| Advantages and disadvantages | Advantages and disadvantages reported for the effort estimation practices |
| Practices used to evaluate the MVP (RQ4) | Name or short description of the practices |
| Advantages and disadvantages | Advantages and disadvantages reported for the evaluation practices |
| Type of research (RQ5) | Classification based on (Wieringa et al., 2006): evaluation research, solution proposal, validation research, philosophical paper, opinion paper, or experience paper. |
| Type of empirical evaluation (RQ6) | Empirical evaluation type: controlled experiment, case study, survey, proof of concept, or none. |

## 5. Focus Group Sessions

To complement the results of the systematic mapping, we organized focus group sessions with industry practitioners to discuss the identified practices. The sessions were conducted based on the guidelines for conducting focus groups in software engineering by Kontio et al. (2008). We also considered recent advice on adapting focus groups to the online environment Menary et al. (2021), given that our focus group was be conducted virtually (during the pandemic).

### *5.1. Participant Selection*

We recruited practitioners involved in the ExACTa[2] (Experimentation-based Agile Co-creation initiative for digital Transformation) initiative at PUC-Rio. This initiative collaborates with several industry partners in innovation projects and currently has about 100 employees. The participant selection was by convenience, as we had direct access to the employees and could invite those that we knew had experience and were involved with developing MVPs. We were able to successfully recruit twelve volunteers to participate in the sessions.

A description of the practitioners' profiles is presented in Table 4. It is possible to note a wide range of profiles. While most of them are working as developers, we were also able to get insights from people working in different positions, such as: UX/UI Leader, Scrum Master, Product Owner, and Research Leader. The majority of them also has a high education level, with ten out of twelve being an MSc or DSc in Computer Science. It is also noteworthy that the range of years of professional experience is wide (2 to 20 years), but that even the least experienced professionals have at least one year working directly with MVPs.

[2]http://www.exacta.inf.puc-rio.br

The recruited participants were involved in projects with a large company that operates in the oil, gas, and energy industry in Brazil. All projects involved outlining and developing MVPs to experiment solution options, followed by MVP increments when suitable. Table 4 also shows the projects in which the participants were involved at ExACTa. The official project names were omitted for customer-related confidentiality reasons. For the same reason, only short descriptions of the purpose of these projects were provided in Table 5. Altogether the participants were involved in developing MVPs in nine different projects. It is noteworthy that the focus was not on details of the projects, but rather on the practices. Nevertheless, all of these projects were innovation projects and used the concept of building MVPs with subsequent increments, which was part of the Lean R&D agile research and development approach in place at ExACTa (Kalinowski et al., 2020).

### *5.2. Session Dynamics*

In a virtual focus group, generally, the number of participants should be reduced Menary et al. (2021) and it is suggested to plan more online focus groups with fewer participants. Following this advice, we decided to split the participants into two focus group sessions. Participants were allocated in these sessions based on their availability.

The two focus group sessions were held remotely via video conference on January 27, 2022 and January 31, 2022, each with six practitioners and two researchers (the first two authors), who acted as facilitators. Participants P1-P6 participated in the first focus group session, whereas participants P7-P12 participated in the second one (*cf.* Table 4).

In order to successfully conduct the focus group activity, the authors had to do some preparation before the sessions. Besides recruiting the participants and allocating them according to their availability, we also prepared a *Participant Characterization Form*. This form had the goal of characterizing each participant in order to better interpret the results. The form consisted of five short questions that allowed us to present the profiles described in Table 4. Finally, we prepared a short presentation containing the practices mapped for each RQ (RQ1-RQ4). These instruments were used during the session as described hereafter.

The sessions followed a three step structure. First, one of the facilitators gave a short introduction of the structure of the session and how it would be conducted. The second step consisted of distributing the *Participant Characterization Form*, via Google Forms, and waiting for all of them to conclude it. As we designed a very quick and simple questionnaire to be answered, we decided to let participants fill it during the activity.

Finally, in the third and last step, the facilitators were responsible to separately introduce each of the research questions (RQ1 to RQ4), allowing each of them to be discussed in isolation. For each of them, facilitators performed two movements. First, the RQ was introduced without mentioning any of the mapped practices, in order to capture unbiased comments from the participants. Then, the facilitator introduced the practices found in the mapping for that RQ, so that the practitioners could talk about their awareness of these practices and their experiences. Each session lasted approximately an hour and a half and both of them were recorded. During the discussion, notes in the form of sticky notes were taken by the researchers and presented to the group in real-time, using the Mural[3] tool.

We used the notes and recorded videos to carefully analyze and summarize the discussions. The analysis of the focus group sessions was conducted by the first author, who wrote summaries of what had been discussed for each research question (similar to the focus group complements included in Section 6). The summaries were peer-reviewed by the second author, who attended the sessions and had access to the videos. As recommended for focus groups Kontio et al. (2008), the results were also shared with the participants, who informally agreed on the overall summarized discussion content.

[3] https://mural.co/

Table 4: Participants Profile

| Id | Academic Background | Years of Experience | Years of Experience with MVPs | Job Title | Projects Involved |
|---|---|---|---|---|---|
| P1 | DSc in Computer Science | 20 | 5 | UX/UI Leader | *All projects of Table 5* |
| P2 | DSc in Computer Science | 5 | 2 | Scrum Master | Project 3, Project 4, Project 5, and Project 8 |
| P3 | MSc in Computer Science | 3 | 1.5 | Developer | Project 2, Project 5, and Project 8 |
| P4 | DSc in Electrical Engineering | 14 | 3 | Developer | Project 5, and Project 8 |
| P5 | MSc in Computer Science | 8 | 2 | Developer | Project 3, Project 5, and Project 8 |
| P6 | MSc in Computer Science | 6 | 3 | Developer | Project 3 and Project 8 |
| P7 | MSc in Computer Science | 4 | 2 | Product Owner | Project 3 and Project 9 |
| P8 | DSc in Computer Science | 8 | 4 | Developer | Project 9 |
| P9 | MSc in Computer Science | 3 | 1 | Developer | Project 9 |
| P10 | BSc in Computer Science | 2 | 1 | Developer | ProA |
| P11 | DSc in Computer Science | 13 | 1.5 | Research Leader | Project 1, Project 5, Project 7, and Project 8 |
| P12 | MSc in Computer Science | 3 | 2 | Developer | Project 3 and Project 9 |

Table 5: Projects description

| Name | Description |
|---|---|
| Project 1 | Comprises developing an intelligent performance control system for chartered ships |
| Project 2 | Concerns developing a novel cognitive safety analysis component for detecting proper usage of personal protective equipment in real-time using data stream from regular closed-circuit television cameras |
| Project 3 | Involves employing artificial intelligence to alert oil refinery's operations and engineering teams about the probability of emissions causing inconvenience to communities |
| Project 4 | Involves using natural language processing to provide information to help refinery inspectors complete reports more quickly and efficiently |
| Project 5 | Involves using artificial intelligence for the prediction of current properties (inferences) in oil refineries, with the objective of increasing the business profitability |
| Project 6 | Involves using process mining to enable an analysis of the process mitigation and resolution of refinery alarms |
| Project 7 | Involves using optimization algorithms to simulate the price of road/rail freight using input parameters |
| Project 8 | Comprises an artificial intelligence solution that continuously processes oil refinery torch burn images to increase the efficiency of burning of gases. |
| Project 9 | Comprises a set of Digital Twin related tools aiming at the integration between the planning, operation, and the Digital Twin |

## 6. Results

This section presents answers to our research questions based on the information extracted from the 33 selected papers as part of our mapping study and complements from our two focus group sessions.

### *6.1. Overview of the selected papers*

The list of selected papers is presented in Appendix A. As Figure 1 shows, the selection process returned papers published between 2013 and 2020. It is possible to observe a concentration of research on the topic in recent years.

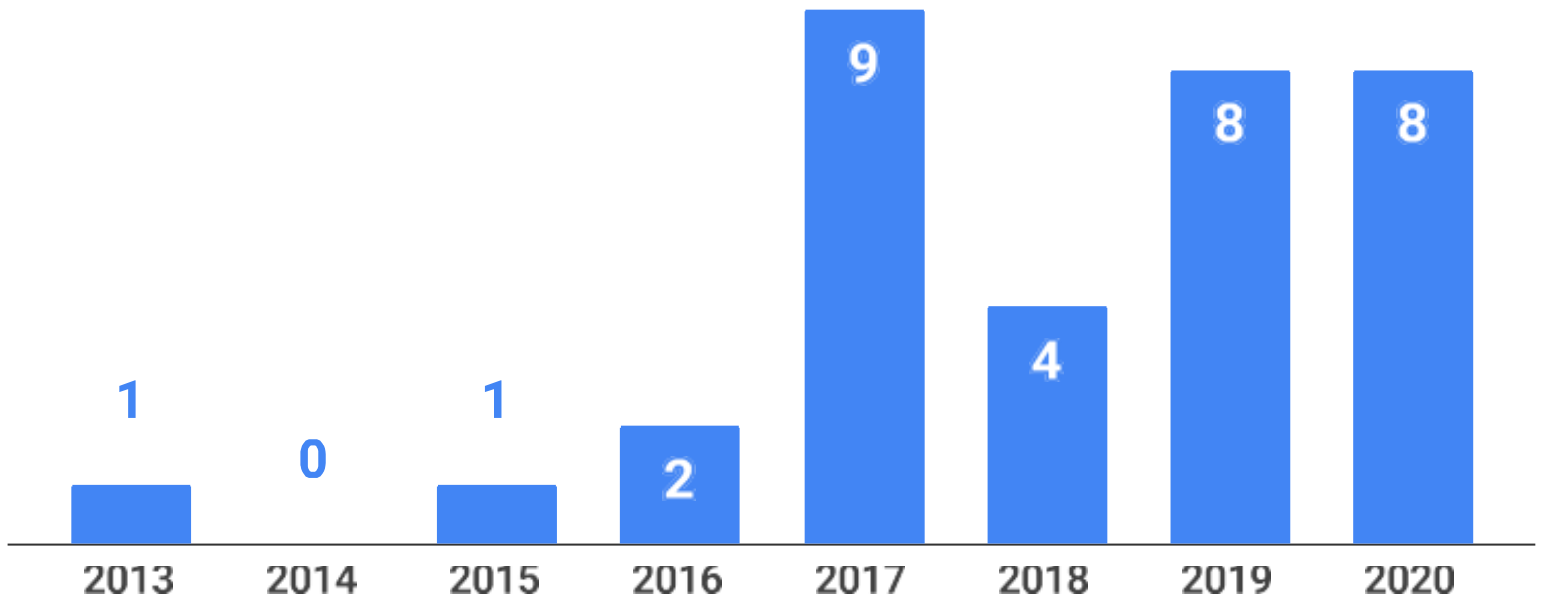

Figure 1: Distribution of Papers per Year

### *6.2. RQ1. Which practices have been used to ideate MVPs?*

#### *6.2.1. Mapping study outcome*

Overall, 73% (24 out of 33) of the papers referenced MVP ideation. Our observations here focus on: approaches; end-user involvement; additional sources of ideas; and business-related, visual prototyping, and documentation practices.

**Approaches.** The adoption of approaches is generally perceived by the use of some of its practices. However, some studies referenced the used approaches directly, such as: Customer Development [S12, S28], Design Thinking [S2, S28], Human-Centered Design [S19], Lean Inception [S5, S33], and User-Centered Design [S3]. For instance, [S2] justifies complementing Lean Startup with Design Thinking, as the former does not provide an approach to ideation. In line with this need, Lean Inceptions were defined as the "combination of Design Thinking and Lean Startup to decide the Minimum Viable Product (MVP)" (Caroli, 2018).

Paper [S25] presented its own ideation approach. It comprises "interdis-ciplinary scenario-building," a series of workshops involving different stakeholders, including end users, business people, and technical people, aiming to create and rank different customer scenarios. Throughout the process, user personas are used to characterize end users. After ranking the scenarios, requirements are derived using a mapping process. The paper reported many advantages related to using these techniques, such as the elucidation of business, user, and technical needs from the outset of the project.

**End-user involvement practices.** End users were reported as being actively involved in the ideation process in 13 studies [S2, S3, S6, S8, S9, S10, S12, S23, S25, S26, S27, S28, S33]. Besides directly involving end-user representatives in workshops [S2, S3, S4, S5, S12, S14, S26, S28, S33], as suggested by the previously mentioned approaches, the most cited method to actively involve end users was through interviews [S2, S6, S8, S10, S23, S26, S27, S28], followed by surveys [S6, S23, S27], and user research [S8, S9, S23]. Moreover, brainstorming sessions [S6, S33], user observation [S2, S27], and focus groups [S26, S30] were also mentioned. For instance, in the rounds of focus groups at the beginning of each development cycle reported in [S30], end users participated alongside many other stakeholders to generate new ideas.

**Additional sources of ideas.** Besides the idea-gathering practices involving end users and multiple stakeholders, some papers [S6, S7, S15, S20, S26] reported the CEO/founders' vision, in a startup context, as the primary source of new product ideas and assumptions about customer problems [S15]. Paper [S26] also reported insights from industry experts. Some structured market research was used in [S1, S2, 27]. Paper [S2] mentioned quantitative and qualitative research on existing solutions in the market and potential users, while paper [S1] described an ideation phase in which the employees' vision and customers' insights were used as inputs for the ideation.

**Business-related practices.** The use of visual business modeling practices was reported in two studies. Paper [S15] presented the use of the Business Model Canvas, and paper [S3] reported the use of Lean Canvas as a way of establishing a shared understanding of the problem, the target audience, and a possible solution.

**Visual prototyping practices.** Some visual prototyping practices were mentioned in the selected studies. For instance, sketching wireframes [S4, S15, S19, S20, S27, S33] and paper prototypes [S22, S23] were mentioned to create a common understanding about the MVP [S22, S23, S27].

**Documentation practices.** The main output of the ideation phase was documented mostly through features (sometimes listed as part of a Canvas, such as an MVP Canvas) and user stories [S3, S4, S11, S12, S19, S26, S27, S28, S33] – general explanation of a software feature written from the perspective of the end user. Sometimes such user stories are complemented by wireframes [S27, S33]. Paper [S27] also presented some additional documentation in the form of customer journeys and stakeholder mappings.

### *6.2.2. Practitioners focus group complement*

While participants mentioned being aware of most of the identified ideation approaches, the discussion regarding **approaches used during MVP ideation** was centered on the Lean Inception workshops (Caroli, 2018), typically performed at the beginning of their projects when following the Lean R&D approach (Kalinowski et al., 2020).

Some lessons learned about the Lean Inception-based ideation were shared. Regarding **end-user involvement**, participant P6 commented that the Project 3 MVP workshop for one oil refinery was attended only by the team, with the end users' vision represented by the Product Owner, resulting in a lot of rework with new alignments during the rest of the MVP development. In contrast, for preparing the Project 3 MVP for another refinery, the workshop was attended by actual end users, which facilitated the understanding of the user's pains and needs. Participants P5 and P7 also commented about the involvement of end users in the workshop, mentioning that sometimes it either does not happen or the opinion of Product Owners and managers has more weight than the opinion of the end users involved. Participant P5 commented on an end user's reaction when asked to evaluate a product idea after the workshop: "I guess you don't know my routine at the refinery. I don't have time to monitor this."

Also, with respect to **scope**, participant P3 mentioned that the scope of the MVP envisioned at the Lean Inception was too broad, leading to a very long list of features. In his opinion, the experts present at the workshop should have acted to reduce the scope of the MVP. He/She mentioned that, in one of the projects he/she participated, a new Lean Inception workshop had to be performed to readjust the scope.

For one of the MVPs, the Design Sprint approach was used. Participant P7, who participated in the activities of this Design Sprint and also in Lean Inceptions, considers the former to be more focused on user-centered design, with the creation of personas and journeys. P7 also believes that, although it does not explore the user as much as a Design Sprint, a Lean

Inception already generates a plan for the development of the MVP. This is reinforced by the vision of P8, who considers Lean Inceptions as a ”cake recipe” with the goal of generating value in the end.

Regarding the **source of ideas**, participant P11 said that the initial idea for a new product usually comes from managers and end users. Participant P1 mentioned that it is also useful to look at the market to analyze similar solutions, and participant P5 reinforced this view by citing the case of Project 2, in which, after several days of discussions, the client presented a market solution with the same purpose as the one being discussed as the scope of the MVP.

About **prototyping practices**, participant P1 made several observations. First, he/she stressed that the creation process should be done with engineers and customers working together after an initial definition of the product scope. In his opinion, the creation of the prototype will serve to “transform words, terms, goals, and plans into images.” In the process described, a wireframe is created together to decrease the uncertainty before developing the high-fidelity prototype. Participant P8 also reinforces the need for end-user involvement in this process since the prototype will serve as a contract between the technical team and the customer. In addition, he/she also cited the importance of the debate that occurs during the creation of the prototype. Participant P1 also mentioned that he/she performed some usability tests to validate the prototypes, but it has been hard to perform them due to the lack of time available.

On **documentation practices**, practitioners cited that it is left up to the artifacts generated during the Lean Inception. Participant P3 noted that he/she consulted these artifacts to understand what had been defined for an MVP, but that he/she had not attended the Lean Inception workshop. After the workshop, everything that should be built is written in the format of User Stories. Along with them are attached the generated prototypes and, eventually, other artifacts, such as a task flow, mentioned by participant P1. Participant P11 considers that User Stories are insufficient for someone outside the team to understand the gathered requirements, but, like P9, he/she believes that they have met the goals of supporting the development of MVPs. Finally, participant P7 sees that, in this context, User Stories are used more as a way to help organizing the tasks than as documentation.

It is possible to observe that the reported ideation practices are well aligned with practices identified as part of the mapping study.

### *6.3. RQ2. Which practices have been used to assess the technical feasibility of MVPs?*

#### *6.3.1. Mapping study outcome*

Only three out of 33 (9%) of our selected papers referred to some kind of technical feasibility assessment. Paper [S25] reported that technical people were involved in ideation workshops to filter technically infeasible scenarios, reducing risks and possible misunderstandings between stakeholders from different backgrounds. It is noteworthy that the involvement of technical stakeholders is also suggested by the aforementioned ideation approaches. More robust assessments are discussed in [S33] and [S14]. Paper [S33] suggests exploring the architecture through a “tracerbullet” strategy (Thomas and Hunt, 2019) to assess whether it is compatible and feasible and that there is a way to solve the problem with reasonable effectiveness, as well as providing a working, demo-able skeleton with some initial implementations. Paper [S14] takes this one step further, reporting a study that explores various software architectures before the MVP development process.

#### *6.3.2. Practitioners focus group complement*

Regarding the technical feasibility analysis of MVPs, some practices were revealed during the focus group. As suggested in paper [S25], there is also the involvement of technical experts in the ideation workshops to signal when they notice something unfeasible being proposed. P1 noted that often the existing technology poses some restrictions, which should be considered during the ideation.

Some actions to guarantee the technical feasibility of the MVP were also conducted after the Lean Inception workshop. Participant P5 explained that the whole team gathers to debate each of the mapped User Stories to perform a risk assessment about it. Participant P2 mentioned that, after ideation, depending on the level of technical uncertainty of the project, experiments are performed so that technical feasibility is evaluated. Participants P5 and P7 also mentioned experiments conducted to test the level of adherence to some of the chosen tools. This care in the choice of tools is due to some bad experiences of options that were not feasible from an economic point of view. In general, participants commented that the teams work with a fail-fast philosophy, doing small experiments, usually informal, to evaluate the technical feasibility before starting the actual development.

Hence, while literature on technical feasibility assessment is rather scarce, the focus group revealed practices such as involving technical staff in the ideation process and conducting informal experiments.

### 6.4. RQ3. Which practices have been used to estimate MVP building effort?

#### 6.4.1. Mapping study outcome

Only two papers directly addressed MVP effort estimation [S29, S31], both focusing on change impact analysis. Paper [S31] evaluates the performance of textual similarity techniques for MVP change impact analysis based on change requests. It uses data from two industrial projects of a Canadian startup. They found that combining textual similarity with file coupling improved impact prediction, and that effort could be predicted with reasonable accuracy (72% to 84%) using textual similarity only. In paper [S29], the same authors, using different methods for textual-similarity analysis, found that the combination of machine-learning techniques with experts' manually added labels had the highest prediction accuracy. According to the authors, better prediction of change impacts allow a company to optimize its resources and provide proper timing of releases for target MVPs.

These papers provide valuable results for change impact analysis. Nevertheless, we found no approach focusing on initial effort estimation for newly outlined MVPs. While the scope of an MVP is supposed to be minimal, typically concerning quick-win implementations, we still see value in its initial effort estimation and identify the need for more research in this context.

#### 6.4.2. Practitioners' focus group complement

In the analyzed context, practitioners reported effort estimates being made at two points in time. Participant P1 mentioned the Lean Inception feature sequencer, used in the ideation phase, as an artifact that gives a sense of effort based on the opinion of more experienced people. Other participants referred to activities that occur during development cycles to estimate the effort to accomplish tasks. Participant P11 mentioned that, for easy tasks, with little uncertainty, the person who will perform the task scores it with story points. For more complex tasks with a high degree of uncertainty, planning poker was performed by the team. Participants P2 and P5 mentioned that, in their projects, the people with more experience in the subject suggest the estimate, which is later validated by the team. Participant P5 also mentioned feature roadmaps being built based on informal estimates and reported that the planning poker practice had been abandoned in recent projects because it was not achieving the expected results. In his opinion, the heterogeneity of the team members made it impossible to make estimates using this method. Furthermore, the fact that the context of innovation projects is very much linked to research makes the degree of uncertainty of several demands very high, and some activities are limited by the available capacity of those responsible for their execution. In summary, we can point out the use of effort estimation activities based on expert opinion and through the group's collective intelligence, in the form of the planning poker practice.

Hence, agile estimation practices such as feature roadmaps and planning poker for user stories, could be employed. However, no such practice has been mentioned in the selected papers. It is also noteworthy that the Lean Inception workshop has a proxy for effort called "waves", onto which sets of features are sequenced. However, there is no contextualized meaning for such "waves" in terms of effort. Possible reasons for the lack of MVP estimation techniques in the literature are explored in the next section.

### *6.5. RQ4. Which practices have been used to evaluate MVPs?*

#### *6.5.1. Mapping study outcome*

Considering our selected papers, 70% (23 out of 33) made some reference to evaluation practices. We discuss them hereafter, semantically grouping them into practices of internal validation, validation with end users, and automated feedback.

**Internal validation practices.** Only one study [S24] reported that the validation was fully performed internally. In this case, the Product Owner (PO) was personally responsible for evaluating the MVP. One of the main advantages reported was that it facilitated a continuous knowledge transfer between the evaluating side (PO) and the development team. Nevertheless, we emphasize the well-known importance of validations involving customers and end-users.

Papers [S14] and [S27] also suggested some form of internal validation. While [S27] reported that the internal validation should happen during internal team meetings, [S14] reported that, during alpha testing and internal review, the team figured out that their product could not be considered an MVP, as it was missing some critical features. Both of them proposed a later external validation with end users.

**Validation with end users.** Validations with end users were conducted in various stages: before the public release of the MVP [S7, S8, S9, S25, S33]; testing on site [S9]; or early on, through mock-ups and early-stage prototypes [S7, S33].

Several evaluation techniques were reported. Paper [S25] reported that workshops with end users were organized to gather feedback for each im-plemented feature before their release, guiding the lean development. Papers [S8], [S11], and [S23] proposed the use of usability tests as controlled experiments. This technique was justified as a way to ensure the MVP's usability, ease of use, ease of learning, and satisfaction ([S11]). Papers [S4, S8] reported the use of Think Aloud Interview Testing before the release of the MVP.

Paper [S33] emphasized the use of continuous experimentation to test MVP-related business hypotheses. Testing these hypotheses involved measuring metrics that reflect changes in business-related results, complemented by end-user questionnaires. Paper [S1] introduced a case that heavily relied on A/B testing and customer interviews, mentioning that the former can lead to conflicts in code and the latter demands a high level of motivation to keep conducting interviews. The A/B testing was also considered hard to extract value when the sample of end users available was small, making it difficult to perform statistical analyses. Even though it recognizes the value of A/B testing as a way to reveal complex knowledge of product usage, paper [S7] justifies its non-use by listing some negative points, such as: the need for a large user base, the use of vast resources to produce variations of the same feature.

Papers [S8], [S11], and [S30] conducted statistical usability tests. Six papers [S1, S2, S12, S16, S21, S32] reported that MVP evaluation was performed by analyzing data generated from the interaction with end users after the release. While paper [S12] stated that market response from the release was analyzed, the others reported more detailed processes, like face-to-face customer interviews [S2] and usage tracking and explicitly collected feed-back [S16, S21]. Paper [S32] analyzed in a case study how pivots can be made in a mobile app based the number of downloads, ratings and reviews.

**Automated Feedback.** Four papers [S13, S16, S18, S21] describe a framework to automate mining of usage tracking and explicit feedback from end users, and reported that the framework allowed to easily capture usage trends.

### *6.5.2. Practitioners focus group complement*

As for the MVP evaluation, for most of the projects (participants explicitly mentioned Project 5, Project 8, and Project 9 in their arguments), the main form of evaluation is to follow up the financial gain generated after the MVP implementation. This follow-up is done by Product Owners and managers using dashboards, being a form of evaluation based on the analysis of data generated after the release. Participant P6 reported that, for some MVPs that rely heavily on Machine Learning algorithms, validation is done from the analysis of data generated from simulations. Participant P1 mentioned that, in Project 4, several MVP usage metrics are automatically collected and presented in a management dashboard and that these quantitative evaluations are appreciated by executives involved in the projects.

Other forms of qualitative validation based on the opinion of the people involved were also commented. Participants P3, P7, and P9 mentioned meetings with end users to perform the validation of MVPs, while participant P5 mentioned validation with the PO during internal review meetings. Participant P7 cited that an interview with end users was conducted as part of the evaluation of the Digital Twin project, and participant P10 mentioned planning to conduct interviews to evaluate Project 4. Participant P5 reported that alpha tests are conducted with people from the team itself to do the evaluation. As a criticism of the evaluation processes adopted, participants also cited the failure to formally evaluate the hypotheses generated during Lean Inception after the MVP development.

Overall, the reported practices were well aligned with the practices identified in literature concerning internal validation practices (*e.g.*, internal review meetings), validations with end-user (*e.g.*, meetings, interviews, and prototype validations), and even automated feedback (*e.g.*, dashboards based on usage data). There were also specific practices identified in literature that participants mentioned being interesting options to address the evaluation that they were not aware of (*e.g.*, statistical usability tests and usage tracking frameworks).

### *6.6. RQ5. Which type of research has been conducted?*

Figure 2 presents the distribution of research types reported in the selected papers. As depicted, *evaluation research* leads with 14 papers, followed by *proposal of solution* and *personal experience*, both with nine papers. The *philosophical paper* category was represented by only one study. This makes sense, as papers reporting industry cases or experiences, where MVPs are widely explored to deliver valuable products, typically involve evaluation research.

### *6.7. RQ6. Which type of empirical evaluations have been performed?*

Figure 3 presents the distribution of empirical evaluations reported. *Case study* evaluations were reported in 13 papers, while three *survey* evaluations and one *proof of concept* were reported. 16 papers did not report any kind of empirical evaluation. It is also worth noticing that no study presented a controlled experiment as an empirical evaluation.

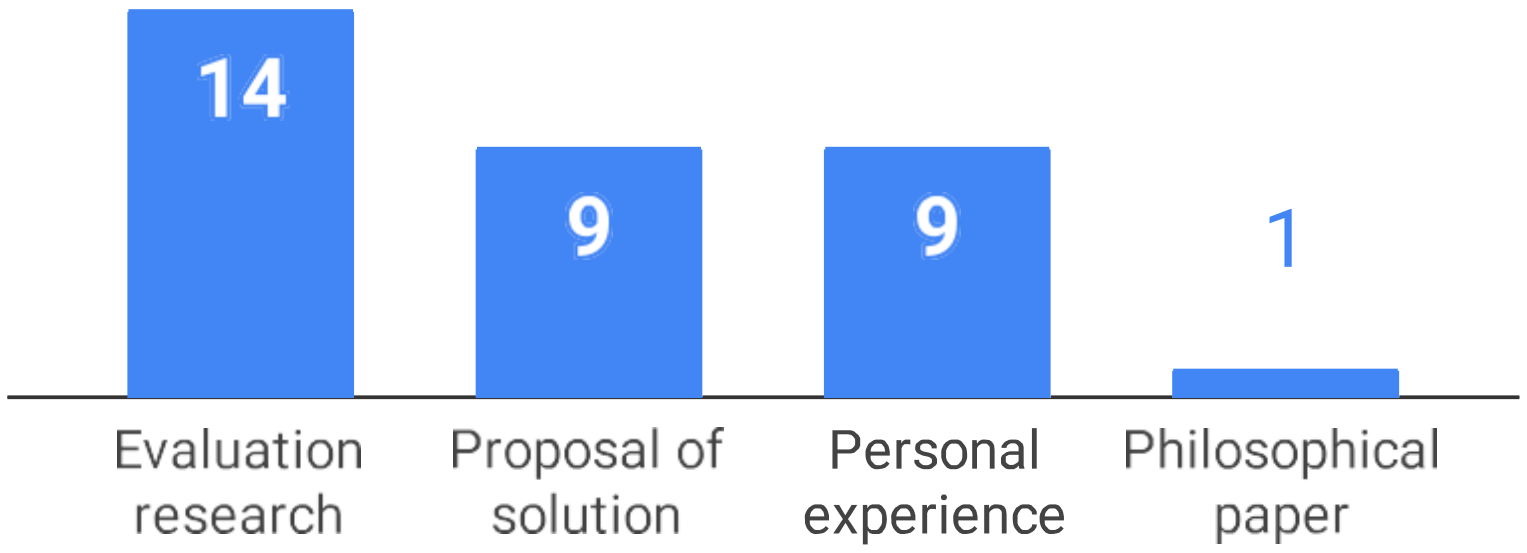

Figure 2: Distribution of Research Type

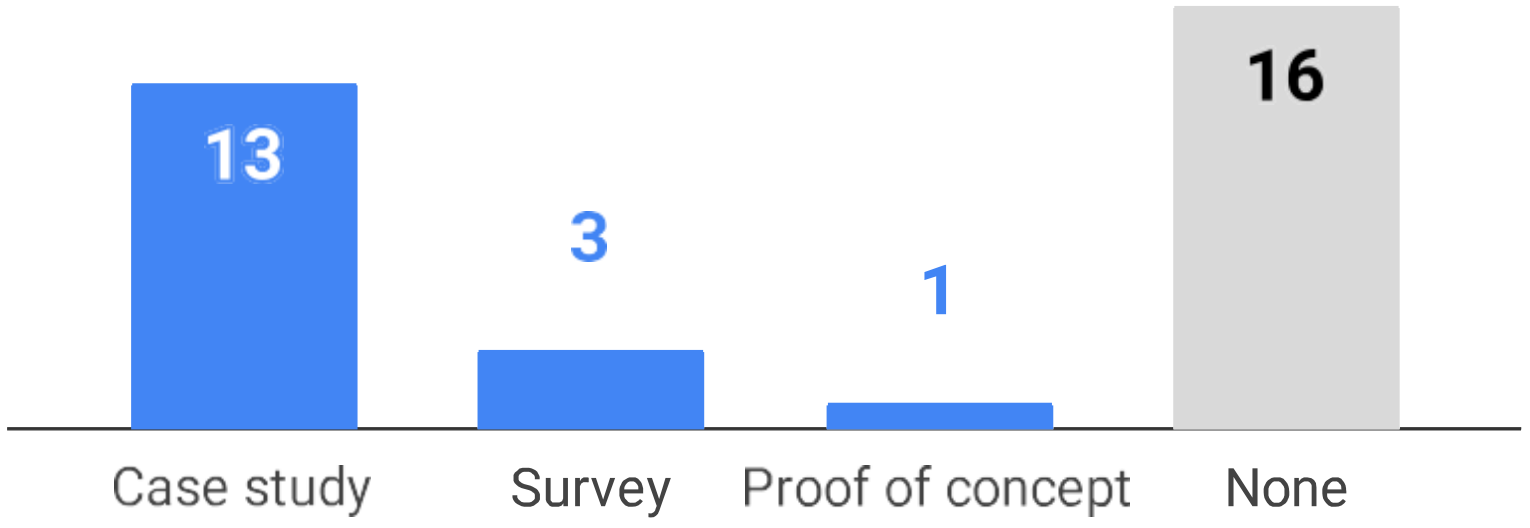

Figure 3: Distribution of Empirical Evaluations

## 7. Discussion

In this section, we summarize our findings and discuss identified gaps and threats to validity.

*7.1. Synthesis of the Results*

The systematic mapping allowed identifying 33 studies concerning the development of software MVPs published by the end of 2020. We analyzed these studies to characterize the practices that have been used to ideate, assess the feasibility, estimate effort, and evaluate such MVPs. Furthermore, we analyzed the types of research and empirical studies conducted. Thereafter we complemented our research from a practitioner's perspective by discussing the mapping study results in two focus group sessions.

Figure 4 shows the mapping of the analyzed MVP activities against their research type and empirical evaluation type. For each of the 33 selected studies, we quantitatively cross-referenced the types of MVP activities (y-axis) cited in each with the type of research and the type of empirical evaluation conducted (x-axis). As an example, study S1 contributed to the MVP activities of MVP ideation and MVP evaluation, had its research type classified as evaluation research, and its empirical evaluation type classified as survey. Thus this study contributed one unit to each of the four point-centered bubbles: Ideation - Evaluation research, Ideation - Survey, Evaluation - Evaluation research, and Evaluation - Survey. This chart is intended to provide us with important insight into how studies of each research type and empirical evaluation are focusing on each type of MVP activity.

It is easy to observe that the identified studies focus on MVP ideation (*cf.* Section 6.2) and evaluation activities (*cf.* Section 6.5). We also noticed that there are many more evaluation research papers than solution proposals. Indeed, many proposals in this area did not emerge from academia (*e.g.*, approaches such as Lean Startup, Lean Inception, and Design Thinking; and practices such as continuous experimentation, A/B testing) and academia has mainly assessed the use of such proposals and their related practices through case studies. The confidence in having representative results concerning the identified practices was improved during the focus group sessions, given that practitioners were aware of many of these practices and had applied some of them within their projects, as described in Section 6. Still, they found the overview comprehensive and mentioned that practices described in some of the identified research could be used to improve their MVP development (*e.g.*, usage tracking frameworks). Their reported experiences also emphasized the importance of some findings (*e.g.*, the importance of involving end users in ideation activities).

The rather scarce research concerning technical feasibility assessment and effort estimation is also worth mentioning. Regarding technical feasibility assessment, this scarcity may be related to the technical simplicity of many MVP projects or to the fact that this activity is probably not receiving enough attention when developing software MVPs. However, not all MVPs are technically simple. In particular, in the digital transformation context, MVPs commonly involve applying new technologies in domains in which they have not been applied before. Failing to properly assess the feasibility may incur wasted investments. Hence, there seems to be a need for solution proposals concerning lightweight software MVP technical feasibility assessment. Indeed, practitioners reported using practices such as involving technical staff in the ideation process and conducting lightweight informal experiments (*e.g.*, to assess if a reasonable estimation accuracy could be obtained in an MVP involving machine learning).

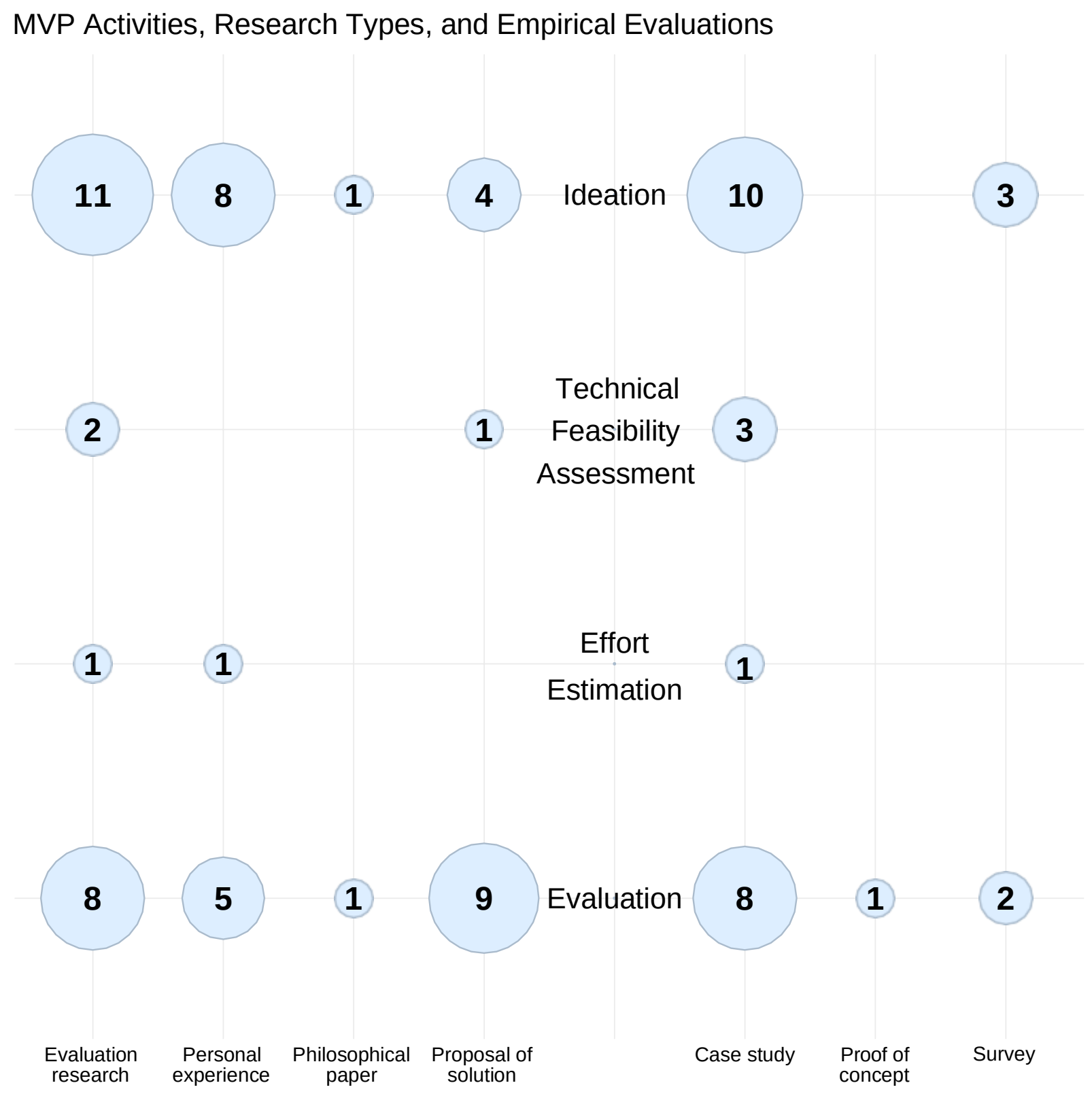

Figure 4: MVP Activities, Research Types, and Empirical Evaluations

With respect to effort estimation, while there is some research on change impact analysis, there is an absence of proposals and evaluations concerning initial MVP effort estimation. This absence may be related to the intended minimal MVP scope. We hypothesize that the limited scope makes stakeholders feel more confident in informal estimates and in informally sketched feature development roadmaps, as reported in our focus group sessions. However, even for MVPs there may be strict time-to-market or cost constraints. Hence, effort estimation should not be neglected and effective techniques should be proposed (or adapted) and evaluated for the MVP context.

Finally, the absence of validation research and controlled experiments within the selected papers is an indication that academia might still be catching up on investigations concerning the MVP topic. Despite the widespread use and importance of MVPs in the software industry – in the startup context and beyond –, this topic still lacks research maturity.

### *7.2. Threats to Validity*

In this section, we critically review our study regarding its threats to validity. We also address the efforts employed to mitigate each of the identified threats.

#### *7.2.1. Internal Validity*

To perform this mapping study, we followed the guidelines proposed by Petersen et al. (2015) and a hybrid search strategy that has been effective in identifying relevant software engineering research (Mourão et al., 2020). While we were not able to find any studies on the topic which we had not included, as in any secondary study, there is a risk of having missed papers. With respect to the focus group, we recruited practitioners that were involved in developing MVPs within nine different projects. Also, as the focus group discussions took place within a predefined format, this could have influenced the level of activity. The moderators avoided interfering in the participants' observations and we noted an overall active participation.

#### *7.2.2. External Validity*

Concerning the generalizability of the results, we invested effort in defining our protocol to enable reproducing our results. Moreover, the intermediate spreadsheets used to control the study selection and the extracted data from all the studies are available and auditable.[1] The fact that the identified practices were understood, mostly recognized, and sometimes even applied by the focus group participants also improved our confidence in having provided a comprehensive and meaningful overview. However, we conducted only two focus group sessions, and they were scoped to a specific context (*e.g.*, all participants were from the ExACTa initiative). Hence, although its results provided valuable practitioner insights, we used them informally to discuss our mapping study findings, without any specific external validity claims on the focus group session results.

#### *7.2.3. Reliability*

The mapping protocol was discussed with other researchers. The study selection and extraction process was conducted by two researchers collaboratively (with one of them validating the activities of the other one). Whenever there was some divergence between the two researchers involved in paper selection and/or data extraction, a third researcher was included in the discussion, so that it was possible to reach a consensus. The overall study selection and data extraction was also reviewed by a third independent researcher to reduce the risk of bias during paper selection and the possibility of errors during data extraction. Also to improve reliability, the selected studies and the extracted data are available online.[1] Unfortunately, we could not make the focus group session video recordings available without compromising the participants' anonymity, but two researchers participated in the focus group sessions and the summaries regarding the discussions held on each mapping study research question were carefully made based on the recorded sticky notes and on the recordings, then peer reviewed and explicitly communi-cated to the participants, who did not raise any objection.

## 8. Conclusion

In this paper, we presented the results of a systematic mapping study on the use of software engineering practices to develop MVPs, complemented by practitioner insights gathered through two focus group sessions. We identified and analyzed 33 papers, published between 2013 and 2020. From these papers, we extracted information related to practices concerning the MVP ideation, technical feasibility assessment, effort estimation, and evaluation. We also analyzed the type of research and the type of empirical evaluation. Furthermore, we conducted two focus group sessions with practitioners extensively involved in the development of MVPs to discuss the practices identified in the mapping study.

Our results show that most papers presented practices regarding MVP ideation and evaluation. For ideation, we found the use of several different approaches (*e.g.*, Customer Development, Design Thinking, Lean Inception), with an emphasis on end-user involvement practices (*e.g.*, workshops, interviews, surveys) and lightweight documentation (*e.g.*, features, user stories, wireframes). Regarding evaluation, there is an emphasis on end-user validations based on different practices (*e.g.*, usability tests, A/B testing, usage data analysis). The awareness of the focus group practitioners of the identified practices, and the alignment with their own practices, as reported in Section 6, improved our confidence in having provided meaningful results.

Another relevant finding is that there is limited research regarding MVP technical feasibility assessment and effort estimation. Our focus group ses-sions raised practices used to carry out these two activities. For instance, on technical feasibility assessment, practitioners reported on the use of informal experiments and on discussions among the development team to perform a risk assessment of user stories. With respect to effort estimation, expert opinion and practices based on collective intelligence, such as planning poker, were cited. Overall, we believe that the focus group sessions allowed to gather valuable insights and lessons learned about the use of such practices, nicely complementing our mapping study results.

Finally, we observed a lack of validation studies and controlled experiments, which points to a limited academia involvement in research on the topic. Indeed, the available evidence does not yet allow to draw precise conclusions on the benefits and limitations of the identified practices, which represents a road still to be paved by future research.

## Appendix A. List of Selected Studies